\documentclass[sort&compress,5p]{elsarticle}

\usepackage{hyperref,bm,amsmath,amsfonts,amssymb}

\journal{Physics Letters B}

\begin{document}

\begin{frontmatter}

\title{Role of particle masses in the magnetic field generation driven
by the parity violating interaction}

\author[a,b,c]{Maxim Dvornikov}

\address[a]{Pushkov Institute of Terrestrial Magnetism, Ionosphere and Radiowave Propagation (IZMIRAN), 142190 Troitsk, Moscow, Russia}

\address[b]{Physics Faculty, National Research Tomsk State University, 36 Lenin
Avenue, 634050 Tomsk, Russia}

\address[c]{II. Institute for Theoretical Physics, University of Hamburg, 149 Luruper Chaussee, D-22761 Hamburg, Germany}

\ead{maxdvo@izmiran.ru}

\begin{abstract}
Recently the new model for the generation of strong large scale magnetic fields in neutron stars, driven by the parity violating interaction, was proposed.
In this model, the magnetic field instability results from the modification of the chiral magnetic effect in presence of the electroweak interaction between
ultrarelativistic electrons and nucleons. In the present work we study how a nonzero mass of charged particles,
which are degenerate relativistic electrons and nonrelativistic protons, influences the generation of the magnetic field in frames of this approach.
For this purpose we calculate the induced electric current of these
charged particles, electroweakly interacting with background neutrons and an
external magnetic field, exactly accounting for the particle mass. This
current is calculated by two methods: using the exact solution of
the Dirac equation for a charged particle in external fields and computing the
polarization operator of a photon in matter composed of background neutrons. We show that the induced current is vanishing in both approaches
leading to the zero contribution of massive particles to the generated
magnetic field. We discuss the implication of our results for the
problem of the magnetic field generation in compact stars.
\end{abstract}

%\begin{keyword}
%\texttt{elsarticle.cls}\sep \LaTeX\sep Elsevier \sep template
%\MSC[2010] 00-01\sep  99-00
%\end{keyword}

\end{frontmatter}

%\linenumbers

The origin of extremely strong magnetic fields $B\gtrsim10^{15}\thinspace\text{G}$
in some neutron stars, called magnetars, is a puzzle for modern physics
and astrophysics. Some models accounting for the generation of such magnetic fields
based on, e.g., the turbulent dynamo and strong fossil fields, are reviewed in
Ref.~\cite{Fer15}. However none of these models adequately describes
all the observed characteristics of magnetars. Recently, several models
for the explanation of magnetic fields in magnetars, involving elementary
particle physics approaches, such as the chiral magnetic effect (CME)~\cite{MirSho15} and
the parity violating electroweak interaction~\cite{Vil80a}, were put forward in
Refs.~\cite{BoyRucSha12,SigLei15}.

In Refs.~\cite{DvoSem15a,DvoSem15b,DvoSem15c,Dvo16} we proposed the model
for the magnetic field generation in magnetars based on the instability
of the magnetic field in matter of a neutron star (NS) composed of
electrons and neutrons interacting by the electroweak forces. We could
predict the growth of the seed magnetic field $B_{0}\sim10^{12}\thinspace\text{G}$,
typical for a pulsar, to values expected in magnetars during the time
intervals comparable with magnetars ages. As shown in Refs.~\cite{DvoSem15c,Dvo16},
the magnetic field growth can be powered by the energy of the thermal
motion of background fermions in the NS matter.
%It should be noted
%that, for the first time, the idea that the parity violating electroweak
%interaction could be important for the generation of strong large
%scale cosmic magnetic fields was discussed in Ref.~\cite{VilLea82}.

In Refs.~\cite{DvoSem15a,DvoSem15b,DvoSem15c,Dvo16} we accounted for the
interaction between ultrarelativistic electrons and nonrelativistic
neutrons, which are both highly degenerate, inside NS. It should be
mentioned that the fact that electrons are ultrarelativistic allowed
us to neglect the electron mass and approximately consider the separate
evolution of right and left chiral components of the election-positron field. Such
an approximation was also used in Refs.~\cite{ChaZhi10,Dvo15} where the
generation of toroidal magnetic fields in NSs was discussed.

Despite an electron in NS is ultrarelativistic, it has a nonzero mass. Any nonzero electron mass will diminish the manifestation of CME.
The helicity flip rate $\Gamma_f$ of relativistic electrons in NS matter was recently computed in Refs.~\cite{Dvo16,GraKapRed15}.
The computed $\Gamma_f \sim m_e^2$, where $m_e$ is the electron mass, mixes the chiral projections of ultrarelativistic electrons
reducing the initial chiral imbalance.
Moreover, as mentioned in Ref.~\cite{Vil80}, any nonzero mass of charged particle can make vanishing the induced anomalous electric current resulting in CME.

The main feature of the model in
Refs.~\cite{DvoSem15a,DvoSem15b,DvoSem15c,Dvo16} is the existence of a
nonzero electric current along the magnetic field direction: $\mathbf{J}=\Pi\mathbf{B}$. Such a current is effective, i.e. it exists only in matter.
As shown in Ref.~\cite{DvoSem14}, the Maxwell equations, modified by adding this current, have an unstable solution leading to the exponential
growth of a seed magnetic field. In the present Letter we will carefully analyze the role the mass of charged particles on the generation of
the induced anomalous current in the presence of the parity violating electroweak interaction. It should be mentioned that, besides ultrarelativistic electrons,
the NS matter should also contain the same amount of degenerate and
nonrelativistic protons for the whole NS to be electrically neutral.
The contribution of these protons to the electric current should be also analyzed. 

In this Letter we shall study the generation of an electric current
of charged particles (electrons or protons) electroweakly interacting with background neutrons under
the influence of an external magnetic field. We shall account
for the particle mass exactly. For instance, as was mentioned above,
unlike electrons, protons are nonrelativistic in NS.  To compute the current
we shall use two methods: the exact solution of the Dirac equation
in external fields~\cite{DvoSem15a,DvoSem15b,Vil80,BalPopStu11}
and the calculation of the antisymmetric contribution to the photon
polarization operator in matter~\cite{BoyRucSha12,DvoSem14}. In
both cases we will show that the induced current along the magnetic
field is vanishing provided a nonzero particle mass is accounted for.
Then we discuss the implication of the obtained result for the generation
of magnetic fields in NS.

We start with the brief discussion of the electroweak interaction between a charged particle, a proton or an electron, and neutrons.
This interaction is accounted for in the elastic forward scattering approximation based
on the Fermi model. We consider the background
matter of NS composed mainly of neutrons, which are taken to be unpolarized
and nonmoving macroscopically. Accounting for both neutral and charged
currents contributions~\cite{MohPal04}, we derive the effective
Lagrangian for the interaction of a test charged particle, described by the
bispinor $\psi$, with this nuclear matter,
\begin{equation}\label{eq:Lint}
  \mathcal{L}_{\mathrm{int}} =  -\bar{\psi}\gamma^{0}
  \left(
    V_{\mathrm{L}}P_{\mathrm{L}} + V_{\mathrm{R}}P_{\mathrm{R}}
  \right)
  \psi,
\end{equation}
where $P_{\mathrm{L,R}}=\left(1\mp\gamma^{5}\right)/2$ are the chiral
projectors, $\gamma^{5} = \mathrm{i}\gamma^{0}\gamma^{1}\gamma^{2}\gamma^{3}$,
and $\gamma^{\mu} = \left( \gamma^{0},\bm{\gamma} \right)$ are the Dirac
matrices. The effective potentials $V_{\mathrm{L,R}}$ in Eq.~(\ref{eq:Lint})
are
\begin{align}\label{eq:VLRe}
  V_{\mathrm{L}} = & \sqrt{2}G_{\mathrm{F}}n_{n}
  \left(
    \frac{1}{2}-\sin^{2}\theta_{\mathrm{W}}
  \right),
  \notag
  \\
  V_{\mathrm{R}} = & -\sqrt{2}G_{\mathrm{F}}n_{n}\sin^{2}\theta_{\mathrm{W}},
\end{align}
for electrons and
\begin{align}\label{eq:VLRp}
  V_{\mathrm{L}} = & \sqrt{2}G_{\mathrm{F}}n_{n}
  \left(
    2|V_{ud}|^{2}+\sin^{2}\theta_{\mathrm{W}}-\frac{1}{2}
  \right),
  \notag
  \\
  V_{\mathrm{R}} = & \sqrt{2}G_{\mathrm{F}}n_{n}\sin^{2}\theta_{\mathrm{W}},
\end{align}
for protons. In Eqs.~(\ref{eq:VLRe}) and~(\ref{eq:VLRp}), $G_{\mathrm{F}}=1.17\times10^{-5}\thinspace\text{GeV}^{-2}$
is the Fermi constant, $n_{n}$ is the neutron density, $V_{ud}\approx0.97$
is the element of the Cabibbo-Kobayashi-Maskawa matrix, and $\sin^{2}\theta_{\mathrm{W}}\approx0.23$
is the Weinberg parameter.

Now we compute the induced electric current with help of the exact solution of the Dirac equation in external fields.
The Dirac equation for a charged particle, accounting for the electroweak interaction
with nuclear matter in Eqs.~(\ref{eq:Lint})-(\ref{eq:VLRp})
under the influence of the external magnetic field $\mathbf{B}=(0,0,B)$,
has the form,
\begin{equation}\label{eq:Direq}
  \left[
    \gamma^{\mu}
    \left(
      \mathrm{i}\partial_{\mu}-eA_{\mu}
    \right) -
    m-\gamma^{0}
    \left(
      V_{\mathrm{L}}P_{\mathrm{L}} +V_{\mathrm{R}}P_{\mathrm{R}}
    \right)
  \right]
  \psi = 0,
\end{equation}
where $A^{\mu}=(0,0,Bx,0)$ is the four vector potential in the Landau
gauge, $e$ is the electric charge ($e<0$ for an electron and $e>0$ for a proton), and $m$ is the particle mass.

We start solving Eq.~\eqref{eq:Direq} for positively charged particles with $e>0$, i.e. for protons.
We separate the variables in Eq.~(\ref{eq:Direq}) in the usual way:
$\psi=\exp ( -\mathrm{i}Et + \mathrm{i}p_{y}y + \mathrm{i}p_{z}z ) \psi _{x}$,
where $\psi_{x}=\psi_{x}(x)$ is the bispinor depending on $x$ coordinate
only. It is convenient to choose the Dirac matrices in the chiral
representation~\cite{ItzZub80}
\begin{gather}
  \gamma^{0} =
  \left(
    \begin{array}{cc}
      0 & -1\\
      -1 & 0
    \end{array}
  \right),
  \quad
  \bm{\gamma} =
  \left(
    \begin{array}{cc}
      0 & \boldsymbol{\sigma}\\
      -\boldsymbol{\sigma} & 0
    \end{array}
  \right),
  \notag
  \\
  \label{eq:chirrep}
  \gamma^{5} =
  \left(
    \begin{array}{cc}
      1 & 0\\
      0 & -1
    \end{array}
  \right),
\end{gather}
where $\boldsymbol{\sigma}$ are the Pauli matrices. The bispinor
$\psi _{x}$ can be also represented using the two component chiral
projections as $\psi_{x}^{\mathrm{T}}=(\xi,\eta)$. On the basis of
Eqs.~(\ref{eq:Direq}) and~(\ref{eq:chirrep}), one gets the equations
for $\xi$ and $\eta$,
% 
% for electrons
%
%\begin{align}\label{eq:xieta}
%  \left(
%    \begin{array}{cc}
%      P_{0}+V_{5}-p_{z} & \mathrm{i}\sqrt{eB}D_{+}\\
%      \mathrm{i}\sqrt{eB}D_{-} & P_{0}+V_{5}+p_{z}
%    \end{array}
%  \right)
%  \xi & =-m\eta,
%  \nonumber
%  \\
%  \left(
%    \begin{array}{cc}
%      P_{0}-V_{5}+p_{z} & -\mathrm{i}\sqrt{eB}D_{+}\\
%      -\mathrm{i}\sqrt{eB}D_{-} & P_{0}-V_{5}-p_{z}
%    \end{array}
%  \right)
%  \eta & =-m\xi,
%\end{align}
%
\begin{align}\label{eq:xieta}
  \left(
    \begin{array}{cc}
      P_{0}+V_{5}-p_{z} & \mathrm{i}\sqrt{eB}D_{-}\\
      \mathrm{i}\sqrt{eB}D_{+} & P_{0}+V_{5}+p_{z}
    \end{array}
  \right)
  \xi & =-m\eta,
  \nonumber
  \\
  \left(
    \begin{array}{cc}
      P_{0}-V_{5}+p_{z} & -\mathrm{i}\sqrt{eB}D_{-}\\
      -\mathrm{i}\sqrt{eB}D_{+} & P_{0}-V_{5}-p_{z}
    \end{array}
  \right)
  \eta & =-m\xi,
\end{align}
where $P_{0}=E-\bar{V}$, $\bar{V}=\left(V_{\mathrm{L}}+V_{\mathrm{R}}\right)/2$,
$V_{5}=\left(V_{\mathrm{L}}-V_{\mathrm{R}}\right)/2$, $D_{\pm}=\partial_{\chi}\pm\chi$,
and $\chi=\sqrt{eB}x-p_{y}/\sqrt{eB}$. Assuming that $\psi _{x}\to0$
at $|x|\to\infty$, we shall look for the solution of Eq.~(\ref{eq:xieta})
in the form,
%
% for electrons
%
%\begin{equation}\label{eq:xietaHerm}
%  \xi =
%  \left(
%    \begin{array}{c}
%      C_{1}u_{\mathrm{n}-1}\\
%      -\mathrm{i}C_{2}u_{\mathrm{n}}
%    \end{array}
%  \right),
%  \quad
%  \eta =
%  \left(
%    \begin{array}{c}
%      C_{3}u_{\mathrm{n}-1}\\
%      -\mathrm{i}C_{4}u_{\mathrm{n}}
%    \end{array}
%  \right),
%\end{equation}
%
\begin{equation}\label{eq:xietaHerm}
  \xi =
  \left(
    \begin{array}{c}
      C_{1}u_{\mathrm{n}}\\
      -\mathrm{i}C_{2}u_{\mathrm{n}-1}
    \end{array}
  \right),
  \quad
  \eta =
  \left(
    \begin{array}{c}
      C_{3}u_{\mathrm{n}}\\
      -\mathrm{i}C_{4}u_{\mathrm{n}-1}
    \end{array}
  \right),
\end{equation}
where $u_{\mathrm{n}}=u_{\mathrm{n}}(\chi)$ is the Hermite function,
$\mathrm{n}=0,1,2,\dotsc$, and $C_{i}$, $i=1,\dots,4$, are the
spin coefficients. The explicit form of $u_{\mathrm{n}}$ can be found, e.g., in Ref.~\cite{DvoSem15a}.

Using the following properties of $u_{\mathrm{n}}$: $D_{+}u_{\mathrm{n}}=\sqrt{2\mathrm{n}}u_{\mathrm{n}-1}$
and $D_{-}u_{\mathrm{n}-1}=-\sqrt{2\mathrm{n}}u_{\mathrm{n}}$, as
well as Eq.~(\ref{eq:xietaHerm}) we get the relations between $C_{i}$,
%
% for electrons
%
%\begin{align}\label{eq:C1234}
%  mC_{1,3} +
%  \left(
%    P_{0}\mp V_{5}\pm p_{z}
%  \right)
%  C_{3,1}\mp\sqrt{2eB\mathrm{n}}C_{4,2} & =0,
%  \nonumber
%  \\
%  mC_{2,4} +
%  \left(
%    P_{0}\mp V_{5}\mp p_{z}
%  \right)
%  C_{4,2}\mp\sqrt{2eB\mathrm{n}}C_{3,1} & =0.
%\end{align}
%
\begin{align}\label{eq:C1234}
  mC_{1,3} +
  \left(
    P_{0}\mp V_{5}\pm p_{z}
  \right)
  C_{3,1}\pm\sqrt{2eB\mathrm{n}}C_{4,2} & =0,
  \nonumber
  \\
  mC_{2,4} +
  \left(
    P_{0}\mp V_{5}\mp p_{z}
  \right)
  C_{4,2}\pm\sqrt{2eB\mathrm{n}}C_{3,1} & =0.
\end{align}
We shall normalize the total proton wave function by the condition
\begin{equation}
  \int\mathrm{d}^{3}x
  \psi^{\dagger}_{\mathrm{n},p_{y},p_{z}}
  \psi_{\mathrm{n}',p_{y}',p_{z}'} =
  \delta_{\mathrm{nn}'}
  \delta
  \left(
    p_{y}-p'_{y}
  \right)
  \delta
  \left(
    p_{z}-p'_{z}
  \right).
\end{equation}
Therefore, $C_{i}$ obey the relation
\begin{equation}\label{eq:sumCi}
  \sum_{i=1}^{4}|C_{i}|^{2}=\frac{1}{(2\pi)^{2}}.
\end{equation}
The energy levels can be obtained from Eq.~(\ref{eq:C1234}) in the
form,
\begin{equation}\label{eq:Enlev}
  \left(
    E-\bar{V}
  \right)^{2} =
  \left(
    E_{0}+sV_{5}
  \right)^{2} +
  m^{2},
\end{equation}
where $E_{0}=\sqrt{2eB\mathrm{n}+p_{z}^{2}}$ is the energy of a massless charged
particle in the constant uniform magnetic field and $s=\pm1$. The
energy levels in Eq.~(\ref{eq:Enlev}) coincide with those found
in Ref.~\cite{BalPopStu11}, where an electron interacting with neutrons and an external magnetic field was considered.
The symmetric gauge for the vector potential $\mathbf{A}$ was used in Ref.~\cite{BalPopStu11}.

First, let us study the case $\mathrm{n}>0$. The explicit form of
$C_{i}$ can be found if we use the following expressions:
%
% for electrons
%
%\begin{align}\label{eq:C1234polop}
%  C_{3,4}
%  \left(
%    P_{0}-V_{5}-sE_{0}
%  \right) +
%  mC_{1,2} & =0,
%  \nonumber
%  \\
%  C_{2,4}
%  \left(
%    sE_{0}-p_{z}
%  \right) -
%  \sqrt{2eB\mathrm{n}}C_{1,3} & =0,
%\end{align}
%
\begin{align}\label{eq:C1234polop}
  C_{3,4}
  \left(
    P_{0}-V_{5}-sE_{0}
  \right) +
  mC_{1,2} & =0,
  \nonumber
  \\
  C_{2,4}
  \left(
    sE_{0}-p_{z}
  \right) +
  \sqrt{2eB\mathrm{n}}C_{1,3} & =0,
\end{align}
which result from Eq.~(\ref{eq:C1234}). Note that Eq.~(\ref{eq:C1234polop})
is a consequence of the existence of the additional spin integral
of Eq.~(\ref{eq:Direq}) found in Ref.~\cite{BalPopStu11}. Using Eqs.~(\ref{eq:sumCi})-(\ref{eq:C1234polop}),
we get $C_{i}$ as
\begin{align}\label{eq:C1234fin}
  \left|
    C_{1}
  \right|^{2} = & \frac{s}{(2\pi)^{2}4E_{0}P_{0}}
  \left(
    sE_{0}-p_{z}
  \right)
  \left(
    P_{0}-V_{5}-sE_{0}
  \right),
  \nonumber
  \\
  \left|
    C_{2}
  \right|^{2} = & \frac{s}{(2\pi)^{2}4E_{0}P_{0}}
  \frac{2eB\mathrm{n}
  \left(
    P_{0}-V_{5}-sE_{0}
  \right)}{
  \left(
    sE_{0}-p_{z}
  \right)},
  \nonumber
  \\
  \left|
    C_{3}
  \right|^{2} = & \frac{s}{(2\pi)^{2}4E_{0}P_{0}}\frac{m^{2}
  \left(
    sE_{0}-p_{z}
  \right)}{
  \left(
    P_{0}-V_{5}-sE_{0}
  \right)},
  \nonumber
  \\
  \left|
    C_{4}
  \right|^{2} = & \frac{s}{(2\pi)^{2}4E_{0}P_{0}}
  \frac{m^{2}2eB\mathrm{n}}{
  \left(
    sE_{0}-p_{z}
  \right)
  \left(
    P_{0}-V_{5}-sE_{0}
  \right)}.
\end{align}
If $\mathrm{n}=0$, we can use Eq.~(\ref{eq:C1234}) directly. In
this case we get that
%
% for electrons, $s=1$, $C_{1}=C_{3}=0$
% $\left(E-\bar{V}\right)^{2}=m^{2}+\left(p_{z}+V_{5}\right)^{2}$
%
$s=-1$, $C_{2}=C_{4}=0$, and
%
% for electrons
%
%\begin{align}\label{eq:C13n0}
%  \left|
%    C_{4}
%  \right|^{2} = & \frac{1}{(2\pi)^{2}}\frac{
%  \left(
%    P_{0}+V_{5}+p_{z}
%  \right)^{2}}{
%  \left(
%    P_{0}+V_{5}+p_{z}
%  \right)^{2}+m^{2}},
%  \nonumber
%  \\
%  \left|
%    C_{2}
%  \right|^{2} = & \frac{1}{(2\pi)^{2}}\frac{m^{2}}{
%  \left(
%    P_{0}+V_{5}+p_{z}
%  \right)^{2}+m^{2}}.
%\end{align}
% 
\begin{align}\label{eq:C13n0}
  \left|
    C_{1}
  \right|^{2} = & \frac{1}{(2\pi)^{2}}\frac{
  \left(
    P_{0}-V_{5}+p_{z}
  \right)^{2}}{
  \left(
    P_{0}-V_{5}+p_{z}
  \right)^{2}+m^{2}},
  \nonumber
  \\
  \left|
    C_{3}
  \right|^{2} = & \frac{1}{(2\pi)^{2}}\frac{m^{2}}{
  \left(
    P_{0}-V_{5}+p_{z}
  \right)^{2}+m^{2}}.
\end{align}
The energy of the lowest Landau level can be found from $\left(E-\bar{V}\right)^{2}=m^{2}+\left(p_{z}-V_{5}\right)^{2}$.

Now, when we have the proton wave function in the explicit form, we are ready to compute the averaged electric current of these
particles along the magnetic field.
It has the form,
\begin{equation}\label{eq:Jgen}
  J_{z} = e
  \sum_{\mathrm{n}=0}^{\infty}
  \sum_{s}
  \int_{-\infty}^{+\infty}\mathrm{d}p_{y}\mathrm{d}p_{z}
  \psi^{\dagger}\gamma^{0}\gamma^{3}\psi f
  \left(
    E-\mu
  \right),
\end{equation}
where $f(E)=\left[\exp(E/T)+1\right]^{-1}$ is the Fermi-Dirac distribution
function, $T$ is the temperature, and $\mu$ is the chemical potential.
Using Eqs.~\eqref{eq:chirrep} and~(\ref{eq:xietaHerm}), we obtain the quantum mechanical average
\begin{align}
  j_{z}^{(\mathrm{n})} = &
  e\int_{-\infty}^{+\infty}\mathrm{d}p_{y}
  \psi^{\dagger}\gamma^{0}\gamma^{3}\psi
  \notag
  \\
  & =
  e^2 B
  \left(
    \left|
      C_{2}
    \right|^{2} +
    \left|
      C_{3}
    \right|^{2} -
    \left|
      C_{1}
    \right|^{2} -
    \left|
      C_{4}
    \right|^{2}
  \right).
\end{align}
On the basis of Eqs.~(\ref{eq:C1234fin}) and~(\ref{eq:C13n0})
we get that
\begin{equation}\label{eq:jzn}
  j_{z}^{(\mathrm{n}>0)} = -\frac{e^2B}{(2\pi)^{2}}\frac{sp_{z}(V_{5}+sE_{0})}{E_{0}P_{0}},
\end{equation}
for $\mathrm{n}>0$, and
%
% for electrons
%
%\begin{equation}\label{eq:jz0}
%  j_{z}^{(\mathrm{n}=0)}=-\frac{e^2B}{(2\pi)^{2}}\frac{p_{z}+V_{5}}{P_{0}},
%\end{equation}
%
\begin{equation}\label{eq:jz0}
  j_{z}^{(0)}=-\frac{e^2B}{(2\pi)^{2}}\frac{p_{z}-V_{5}}{P_{0}},
\end{equation}
for $\mathrm{n}=0$. Using Eqs.~\eqref{eq:Enlev}, (\ref{eq:jzn}) and~(\ref{eq:jz0}),
one obtains that after the statistical averaging, 
\begin{align}\label{eq:intpz}
  J_z = &
  \sum_{\mathrm{n}=0}^{\infty}
  \sum_{s}
  \langle
    j_{z}^{(\mathrm{n})}
  \rangle=0,
  \notag
  \\
  \langle
    j_{z}^{(\mathrm{n})}
  \rangle = &
  \int_{-\infty}^{+\infty}\mathrm{d}p_{z}j_{z}^{(\mathrm{n})} f
  \left(
    E-\mu
  \right) = 0,
\end{align}
for any $\mathrm{n}$.

The fact that $\langle j_{z}^{(\mathrm{n}>0)}\rangle = 0$ is obvious. Indeed one can see in Eq.~\eqref{eq:Enlev} that, at $\mathrm{n}>0$,
both $P_0$ and $E_0$ are even in $p_z$
making $j_{z}^{(\mathrm{n}>0)}$ in Eq.~\eqref{eq:jzn} odd in $p_z$. Thus the integration over $p_z$ in Eq.~\eqref{eq:intpz} gives
$\langle j_{z}^{(\mathrm{n}>0)}\rangle = 0$.
To demonstrate that $\langle j_{z}^{(0)}\rangle = 0$ we recall that, at $\mathrm{n}=0$,
$E=\bar{V} + \sqrt{m^{2}+\left(p_{z}-V_{5}\right)^{2}}$ for particles. Then, changing the integration variable $p_{z} \to p_{z}' = p_{z}-V_{5}$,
one obtains that $j_{z}^{(0)}$ in Eq.~\eqref{eq:jz0} is odd in $p_{z}'$.
Integrating over $p_{z}'$ in Eq.~\eqref{eq:intpz} from $-\infty$ to $+\infty$ (see below), one gets that $\langle j_{z}^{(0)}\rangle = 0$. 
Restoring vector notations in Eq.~\eqref{eq:intpz}, we obtain that the electric current along magnetic field is vanishing: $\mathbf{J} =\Pi \mathbf{B} = 0$.
Analogously one can show the absence of the contribution of massive antiparticles
to this electric current.

One can demonstrate that the induced anomalous current of electrons along the magnetic field is also vanishing.
For this purpose one should either find the electron wave function in the Landau gauge analogously to Eqs.~\eqref{eq:xieta}-\eqref{eq:C13n0},
and then compute the current as in Eqs.~\eqref{eq:Jgen}-\eqref{eq:intpz}; or use the wave function of an electron,
interacting with background neutrons under the influence of the magnetic field, found in Ref.~\cite{BalPopStu11} in the symmetric gauge.
We shall omit these computations for brevity.

The reason for the disappearance of the electric current for massive particles is the following. It is well known that, in case of massless particles,
the nonzero
electric current along the external magnetic field is due to the polarization effects of charged particles
at zero Landau level~\cite{Vil80}. It is actually the manifestation of CME~\cite{MirSho15}. The momentum of massless particles
is correlated with the particle spin. The particle spin, in its turn, is correlated with the magnetic field direction at $\mathrm{n}=0$.
Therefore, for massless charged particles at zero Landau level, the particle momentum will have a certain direction with respect to the magnetic field,
i.e. $p_z$ will vary either from $0$ to $+\infty$ or from $-\infty$ to $0$ depending on the particle charge~\cite{DvoSem15a,DvoSem15b}.
Therefore, if we consider the analogue of Eq.~(\ref{eq:intpz}) for massless
particles, the integration over $p_z$ will give a nonzero result. On the contrary, for massive particles, $p_z$ is no longer correlated with the magnetic field,
changing from $-\infty$ to $+\infty$. It happens even at $\mathrm{n}=0$ and makes $J_{z}$ to vanish.

We can also demonstrate the cancellation of the induced current along
the magnetic field direction in case of massive particles using the
results of the one loop calculation of the polarization operator $\Pi_{\mu\nu}$ in
Ref.~\cite{DvoSem14}. A nonzero antisymmetric part $\Pi_{ij}=\mathrm{i}\varepsilon_{ijn}k^{n}\Pi$
of the polarization operator can induce the current along the magnetic
field: $J^{i}=-\Pi_{ij}A^{j}=\Pi B_{i}$ or $\mathbf{J}=\Pi \mathbf{B}$. Performing analogous one loop computation
of the polarization operator of a photon in a medium composed of electrons, protons
and neutrons as in Ref.~\cite{DvoSem14}, one gets the new form factor
$\Pi$ in the limit $k^{2}\ll m^{2}$ as
%
%\begin{widetext}
\begin{align}\label{eq:Pi2pn}
  \Pi & = -\frac{7}{3}e^{2}V_{5}
  \int\frac{\mathrm{d}^{3}p}{(2\pi)^{3}}\frac{1}{\mathcal{E}_{\mathbf{p}}^{3}}
  \nonumber
  \\
  &
  \times
  \biggl\{
    \frac{m^{2}}{\mathcal{E}_{\mathbf{p}}^{2}}
    \left[
      \frac{1}{\exp[\beta(\mathcal{E}_{\mathbf{p}}-\mu)]+1} +
      \frac{1}{\exp[\beta(\mathcal{E}_{\mathbf{p}}+\mu)]+1}
    \right]
    \nonumber
    \\
    & +
    \frac{m^{2}\beta}{2\mathcal{E}_{\mathbf{p}}}
    \left[
      \frac{1}{\cosh[\beta(\mathcal{E}_{\mathbf{p}}-\mu)]+1} +
      \frac{1}{\cosh[\beta(\mathcal{E}_{\mathbf{p}}+\mu)]+1}
    \right]
    \nonumber
    \\
    & -
    \frac{\beta^{2}\mathbf{p}^{2}}{6}
    \left[
      \frac{\tanh[\beta(\mathcal{E}_{\mathbf{p}}-\mu)/2]}
      {\cosh[\beta(\mathcal{E}_{\mathbf{p}}-\mu)]+1} +
      \frac{\tanh[\beta(\mathcal{E}_{\mathbf{p}}+\mu)/2]}
      {\cosh[\beta(\mathcal{E}_{\mathbf{p}}+\mu)]+1}
    \right]
  \biggr\} ,
\end{align}
%\end{widetext}
%
where $\mathcal{E}_{\mathbf{p}}=\sqrt{\mathbf{p}^{2}+m^{2}}$, $\beta=1/T$
is the reciprocal temperature, and $k^{\mu}$ is the photon momentum. In Eq.~\eqref{eq:Pi2pn} we consider neutrons as background fermions.

In Refs.~\cite{DvoSem15a,DvoSem15b,DvoSem15c,Dvo16}, we are mainly interested in the generation of magnetic field in NS, when it is in a thermal equilibrium,
which is reached after $\sim 10^2\,\text{yr}$ after the supernova collapse. At this stage of the NS evolution, charged particles, i.e. electrons and
protons, are highly degenerate. Hence we should consider the limit $\mu/T\gg1$
in Eq.~(\ref{eq:Pi2pn}). Taking into account the identities,
\begin{align}
  \lim_{\beta\to\infty}
  \frac{\beta}{\cosh(\beta x)+1} = &
  2\delta(x),
  \notag
  \\
  \lim_{\beta\to\infty}
  \frac{\beta^{2}\tanh(\beta x/2)}{\cosh(\beta x)+1} = &
  -2\delta'(x),
\end{align}
which were derived in Ref.~\cite{DvoSem14}, and recalling that for
degenerate fermions one has $\mu_{e,p}=\sqrt{\left(3\pi^{2}n_{e,p}\right)^{2/3}+m^{2}}>0$,
where $n_{e,p}$ are the densities of electrons and protons, one obtains that $\Pi=0$
in Eq.~(\ref{eq:Pi2pn}). Thus we again get that there is no electric
current of massive charged particles along the magnetic field.

Note that, the main reason to get $\Pi=0$ in Eq.~\eqref{eq:Pi2pn} is to consider $k^\mu = 0$ in the computation of the polarization tensor.
It corresponds to the zero momentum of a photon/plasmon in NS matter or the static external magnetic field.
It is worth to mention that, in contrast to the present work, in Ref.~\cite{DvoSem14},
we assumed that $k^2 = \omega_p^2>0$, where $\omega_p$ is the plasma frequency, i.e. we considered the propagation of an electromagnetic wave there.

In conclusion we mention that we have shown that the induced electric
current along the external magnetic field of massive charged particles, electroweakly interacting with background neutrons, is vanishing.
We have studied one particular
implementation of this problem: the parity violating electroweak interaction in the Fermi approximation; cf. Eqs.~\eqref{eq:Lint}-\eqref{eq:VLRp}.
We have demonstrated the current cancellation using two methods: the exact
solution of the Dirac equation in external fields and the analysis
of the photon polarization operator. The former approach is beyond
the perturbation theory whereas, in the later method, one demonstrates
the washing out of the current linear in $G_{\mathrm{F}}$ and the
fine structure constant $\alpha_{\mathrm{em}}=e^{2}/4\pi$.

Note that, for the first time, the cancellation of the induced current of electroweakly interacting
massive particles was mentioned in Ref.~\cite{Vil80}, whereas
for massless particles a nonzero current may well exist~\cite{DvoSem15a,DvoSem15b,Vil80}.
However, this observation was made in Ref.~\cite{Vil80} on the basis of the perturbative computation of the one loop contribution to the photon polarization
tensor. The novelty of the present work compared to the result of Ref.~\cite{Vil80},
is that we have demonstrated the disappearance of the current using the exact solution of the Dirac equation in all orders
in $G_{\mathrm{F}}$ and $\alpha_{\mathrm{em}}$, i.e. nonperturbatively.

Such an unusual
dependence of the induced current on the charged particle masses is
related to the breaking of the chiral symmetry for massive particles. Massive and massless particles belong
to different phases in which the chiral symmetry is broken and restored. The restoration of the chiral symmetry can take place in
the presence of background matter having high temperature and/or density. The size of ``bubbles'', containing matter in the symmetric phase, will depend
smoothly on the temperature $T$ and/or the density $\rho$ of background matter. The nonzero anomalous
current $\mathbf{J} = \Pi \mathbf{B}$,
which results in the magnetic field instability,
will exist only in ``bubbles'' with restored chiral symmetry. Therefore, if one studies the generation of a magnetic field driven by CME
in a realistic cosmological/astrophysical media accounting for the chiral phase transition, the scale and the strength of
this magnetic field will be smooth functions of $T$ and/or $\rho$.

It should be noted that, at the absence of the electroweak interaction, the disappearance of CME~\cite{Vil80},
i.e. the cancellation of the induced current $\mathbf{J} = 2(\alpha_\mathrm{em}/\pi) \mu_5 \mathbf{B} = 0$ for massive particles can be foreseen.
Here $\mu_5 = (\mu_\mathrm{R} - \mu_\mathrm{L})/2$ and $\mu_\mathrm{R,L}$ are the chemical potentials of right and left particles.
Indeed, if $m\neq 0$, the decomposition to the left and right chiral projections is impossible and we should set $\mu_5 = 0$
since for massive particles there should be only one chemical potential $\mu = \mu_\mathrm{R} = \mu_\mathrm{L}$.
However, if the electroweak interaction with background fermions is present, the induced anomalous current for massless particles was found in
Refs.~\cite{DvoSem15a,DvoSem15b} in the form $\mathbf{J} = 2(\alpha_\mathrm{em}/\pi) \left( \mu_5 + V_5 \right) \mathbf{B}$.
The washing out of this current for massive particles is not obvious since $V_\mathrm{L} \neq 0$ and $V_\mathrm{R} \neq 0$
in Eqs.~\eqref{eq:Lint}-\eqref{eq:VLRp}, giving one $V_5 \neq 0$ for both massless and massive particles.
Thus the demonstration that CME is vanishing for massive particles in the presence of the electroweak interaction requires a special analysis which,
in fact, was carried out in the present Letter.

The results of our
work are equally applied for the currents of massive electrons and protons. As mentioned above, despite electrons
are ultrarelativistic in NS they possess nonzero masses. Therefore, basing on our results, the generation of magnetic fields in magnetars
driven by the electron-nucleon interaction, proposed in Refs.~\cite{DvoSem15a,DvoSem15b,DvoSem15c,Dvo16}, is questionable unless
there is a mechanism restoring the chiral symmetry for electrons in
NS. As found in Ref.~\cite{Rub86}, the electroweak phase transition
in dense matter can happen if the matter density exceeds $n_{\mathrm{cr}}\sim M_{\mathrm{W}}^{3}\approx6.6\times10^{46}\thinspace\text{cm}^{-3}$,
where $M_{\mathrm{W}}\approx80\thinspace\text{GeV}$ is the W-boson
mass. This value is far beyond the density in NS. The same disappointing arguments are valid with to respect to the findings of Refs.~\cite{SigLei15,ChaZhi10}.

Nevertheless we can still use the approach of Refs.~\cite{DvoSem15a,DvoSem15b,DvoSem15c,Dvo16}
for the generation of magnetic fields in compact stars considering the
quark-quark electroweak interaction~\cite{MohPal04}. The chiral symmetry was shown
in Ref.~\cite{DexSch10} to be restored for the lightest $u$ and
$d$ quarks for a specific equation of state of nuclear matter in a hybrid star, i.e. in
NS having quark matter core, or in a hypothetical quark star~\cite{Gle00}. Moreover, as shown in Ref.~\cite{Fae97},
the effective masses of baryons can be significantly reduced if QCD radiative corrections are taken into account.
It is the indication to the fact that the chiral symmetry can be restored in dense matter. The results of Refs.~\cite{DvoSem15a,DvoSem15b,DvoSem15c,Dvo16}
can be straightforwardly applied to describe the magnetic field generation in a quark/hybrid star. The consideration of the details of the magnetic
field generation driven by the electroweak quark-quark interaction will be done in our forthcoming work.

\section*{Acknowledgements}

I am thankful to A.A.~Andrianov, A.V.~Borisov, V.V. Bra\-gu\-ta, M.I.~Krivoruchenko, A.E.~Lobanov, B.V.~Martemyanov, G.~Sigl, M.I.~Vysotsky, V.I.~Zakharov,
and V.Ch. Zhukovsky for useful discussions as well as to the Competitiveness Improvement Program at the Tomsk State University,
RFBR (research project No.~15-02-00293), and DAAD (grant No. 91610946) for partial support.


\begin{thebibliography}{40}

\bibitem{Fer15}
  L.~Ferrario, A.~Melatos, J.~Zrake,
  Magnetic field generation in stars,
  Space Sci. Rev. 191 (2015) 77--109,
% doi:10.1007/s11214-015-0138-y
  arXiv:1504.08074~[astro-ph.SR].

\bibitem{MirSho15}
  V.A.~Miransky, I.A.~Shovkovy,
  Quantum field theory in a magnetic field:
  From quantum chromodynamics to graphene and Dirac semimetals,
  Phys. Rept. 576 (2015) 1--209,
  arXiv:1503.00732~[hep-ph].

%\bibitem{AkaYam13}
%  Y.~Akamatsu, N.~Yamamoto,
%  Phys. Rev. Lett. 111 (2013) 052002,
%  arXiv:1302.2125~[nucl-th].

\bibitem{Vil80a}
  A.~Vilenkin,
  Cancellation of equilibrium parity violating currents,
  Phys. Rev. D 22 (1980) 3067--3079.

\bibitem{BoyRucSha12}
  A.~Boyarsky, O.~Ruchayskiy, M.~Shaposhnikov,
  Long-range magnetic fields in the ground state of
  the Standard Model plasma,
  Phys. Rev. Lett. 109 (2012) 111602,
  arXiv:1204.3604~[hep-ph].

\bibitem{SigLei15}
  G.~Sigl, N.~Leite,
  Chiral magnetic effect in protoneutron stars and
  magnetic field spectral evolution,
  J. Cosmol. Astropart. Phys. 01 (2016) 025,
  arXiv:1507.04983~[astro-ph.HE].

\bibitem{DvoSem15a}
  M.~Dvornikov, V.B.~Semikoz,
  Magnetic field instability in a neutron star driven by
  the electroweak electron-nucleon interaction versus
  the chiral magnetic effect,
  Phys. Rev. D 91 (2015) 061301,
  arXiv:1410.6676~[astro-ph.HE].

\bibitem{DvoSem15b}
  M.~Dvornikov, V.B.~Semikoz,
  Generation of the magnetic helicity in a neutron star driven by
  the electroweak electron-nucleon interaction,
  J. Cosmol. Astropart. Phys. 05 (2015) 032,
  arXiv:1503.04162~[astro-ph.HE].

\bibitem{DvoSem15c}
  M.~Dvornikov, V.B.~Semikoz,
  Energy source for the magnetic field growth in magnetars driven by
  the electron-nucleon interaction,
  Phys. Rev. D 92 (2015) 083007,
  arXiv:1507.03948~[astro-ph.HE].

\bibitem{Dvo16}
  M.~Dvornikov,
  Chiral imbalance evolution in dense matter and the generation of
  magnetic fields in magnetars,
  %Phys. Rev. D 00 (2016) 000000,
  arXiv:1510.06228~[hep-ph].

%\bibitem{VilLea82}
%  A.~Vilenkin, D.A.~Leahy,
%  Astrophys. J. 254 (1982) 77--81.

\bibitem{ChaZhi10}
  J.~Charbonneau, A.~Zhitnitsky,
  Topological currents in neutron stars:
  Kicks, precession, toroidal fields, and magnetic helicity,
  J. Cosmol. Astropart. Phys. 08 (2010) 010,
  arXiv:0903.4450~[astro-ph.HE].

\bibitem{Dvo15}
  M.~Dvornikov,
  Galvano-rotational effect induced by electroweak interactions in pulsars,
  J. Cosmol. Astropart. Phys. 05 (2015) 037,
  arXiv:1503.00608~[hep-ph].

\bibitem{Vil80}
  A.~Vilenkin,
  Equilibrium parity violating current in a magnetic field,
  Phys. Rev. D 22 (1980) 3080--3084.

\bibitem{GraKapRed15}
  D.~Grabowska, D.~B.~Kaplan, and S.~Reddy,
  Role of the electron mass in damping chiral plasma instability
  in supernovae and neutron stars,
  Phys. Rev. D 92 (2015) 085035,
  arXiv:1409.3602~[hep-ph].

\bibitem{DvoSem14}
  M.~Dvornikov, V.B.~Semikoz,
  Instability of magnetic fields in electroweak plasma driven by
  neutrino asymmetries,
  J. Cosmol. Astropart. Phys. 05 (2014) 002,
  arXiv:1311.5267~[hep-ph].

\bibitem{BalPopStu11}
  I.A.~Balantsev, Yu.V.~Popov, A.I.~Studenikin,
  On the problem of relativistic particles motion in
  strong magnetic field and dense matter,
  J. Phys. A 44 (2011) 255301,
  arXiv:1012.5592~[hep-ph].

\bibitem{MohPal04}
  R.N.~Mohapatra, P.B.~Pal,
  Massive Neutrinos in Physics and Astrophysics, third ed.,
  World Scientific, Singapore, 2004, pp.~5--8.

\bibitem{ItzZub80}
  C.~Itzykson, J.-B.~Zuber,
  Quantum Field Theory,
  McGraw-Hill, New York, 1980, pp.~691--696.

\bibitem{Rub86}
  V.A.~Rubakov,
  On the electroweak theory at high fermion density,
  Prog. Theor. Phys. 75 (1986) 366--385.

\bibitem{DexSch10}
  V.~Dexheimer, S.~Schramm,
  A novel approach to model hybrid stars,
  Phys. Rev. C 81 (2010) 045201,
  arXiv:0901.1748~[astro-ph.SR].

\bibitem{Gle00}
  N.K.~Glendenning,
  Compact Stars:
  Nuclear Physics, Particle Physics, and General Relativity, second ed.,
  Springer, New York, 2000, pp.~322--365 and~414--440.

\bibitem{Fae97}
  A.~Faessler, A.J.~Buchmann, M.I.~Krivoruchenko, B.V.~Martemyanov,
  Nuclear matter with a Bose condensate of dibaryons in relativistic
  mean-field theory,
  Phys. Lett. B 391 (2012) 255--260,
  nucl-th/9611020.

\end{thebibliography}
\end{document}